\documentclass[english,aps,twocolumn, floatfix,showpacs,pra]{revtex4-1}
\usepackage[T1]{fontenc}
\usepackage[utf8]{inputenc}
\usepackage{units}
\usepackage{mathtools}
\usepackage{amsmath}
\usepackage{amssymb}
\usepackage{cancel}
\usepackage{stackrel}
\usepackage{graphicx}
\usepackage{color,soulutf8, ulem}
\usepackage{hyperref}
\usepackage{braket}
\usepackage{titlesec}

\DeclareMathOperator{\Tr}{Tr}
\makeatletter
\newcommand{\be}{\begin{equation}}
\newcommand{\ee}{\end{equation}}

\newcommand{\Sx}{\hat{S}_x}
\newcommand{\Sy}{\hat{S}_y}
\newcommand{\Sz}{\hat{S}_z}

\newcommand{\bb}[1]{\left( #1\right)}

\newcommand{\kp}{}%{\color{red}}
\newcommand{\ks}{}%{\color{red}}

\newcommand{\ft}[1]{\footnote{#1}}%{{\footnotesize(#1)}}%{\color{red}}

\makeatother

\titleformat{\section}[runin]{\normalfont\normalsize\bfseries\itshape}{\thesection}{1em}{}

\titleformat{\subsection}[runin]{\normalfont\normalsize\itshape}{\thesubsection}{1em}{}
\usepackage{babel}

\begin{document}

\title{Quantum correlations protection through spin self-rephasing in 1-D Bose gas}
%TODO title

%\author{Konrad Szyma\'{n}ski, Krzysztof Paw\l owski}

\author{Konrad Szyma\'{n}ski}
\affiliation{Center for Theoretical Physics PAS, Aleja Lotnik\'ow 32/46, 02-668 Warszawa, Poland\\
Marian Smoluchowski Institute of Physics, ul. Łojasiewicza 11, 30-348 Kraków, Poland}

\author{Krzysztof Paw\l owski}
\affiliation{Center for Theoretical Physics PAS, Aleja Lotnik\'ow 32/46, 02-668 Warszawa, Poland}
\date{\today}

\begin{abstract} %na razie wziete z plakatow
System consisting of a number of trapped atoms evolving under the influence of
external inhomogenous magnetic field undergoes spin dephasing: classically, since each atom feels different field along its trajectory, the
spin rotation rates differ; as a result the average spin decays. In a quantum mechanical context this
corresponds to entanglement of spin and spatial degrees of freedom and nontrivial internal spin dynamics. The spin dephasing can be prevented by tuning
the interaction between the atoms. This phenomenon, called spin self-rephasing, has been observed
experimentally %\cite{deutsch2010spin}
 and can increase the coherence time by a large factor. While such systems
have been studied from a semiclassical point of view, a quantum mechanical description is limited, {especially in the case of entangled states}. %In this work we fill in the gap by providing a numerical simulation of the behavior of the
%quantum mechanical system of several interacting bosons in the presence of
%inhomogenous magnetic field. 
{\ks In this work we provide a numerical simulation of an ab initio model and provide realistic examples of spin self-rephasing used to counteract the effect of inhomogenous magnetic field.}
We analyze in particular the joint effect of magnetic field inhomogeneity and interactions on the coherent and spin squeezed states evolution. % and analysis of the spin rephasing and coherence.
\end{abstract}
%TODO pacy do sprawdzenia

\maketitle

One of the fundamental limitations of precise measurements is noise introduced by the measurement statistics: if one tries to estimate expectation value of a classical random variable $\phi$, the variance of an estimator diminishes with number of independent experiments $N$ as square root, $\Delta^2 \langle \phi\rangle \propto N^{-1/2}$. 
The spectacular success of the quantum information is utilization of entangled states to beat this limit -- unlike the classical case, quantum correlations between parts (classically: different runs) can decrease the variance \cite{petz2011introduction}. %One of the most useful and comprehensively studied states is so-called squeezed state. 
The gravitational wave detectors {\ks soon are going to } utilize the squeezed state of light to capture the tiny signal coming from the collision of the massive astronomical objects; the squeezed states were also used in the prototypes of the magnetometers, capable to scan the magnetic field fluctuations varying in the micrometer scales \cite{Ockeloen2013}. 

In general, quantum correlations increase the precision of devices measuring unknown parameter $\phi$ of total system Hamiltonian (e.g. magnetic field strength) -- this value can be inferred from experiment outcomes (e.g. realizations of $S_x$ spin projection measurements)
with lower uncertainty: for $N$ subsystems, like qubits or atoms, $\Delta^2 \langle \phi\rangle \propto N^{-1}$ -- so-called Heisenberg limit \cite{giovannetti2011advances}. %  Such precision demands special entangled states, which beats by the factor $N^{1/2}$ the uncorrelated sources. 
Squeezed state is not the only example: one can saturate the Heisenberg limit initiating the particles in the interferometer in a Schr\"odinger cat state \cite{Leibfried2004}. 

However, the strongly correlated states are very fragile and due to their sensitivity to external disturbances, even small amount of noise can completely destroy the quantum correlations, making their applications questionable. In this respect there are very discouraging theoretical results showing that  in the presence of any noise, the precision scales asymptotically like $1/\sqrt{N}$ with at most constant factor improvement \cite{DemkowiczDobrzanski2012Sep}. 
Therefore the huge benefits of entangled states can be  reached only in non-asymptotic limit, for short-time interferometers and when we are restricted to a fixed number of qubits. The larger noises, the shorter the duration at which the highly entangled state can be of importance. 
Here we search for improvements by studying the microscopic, {\it ab initio} model with decoherence to benefit from usually  not-accounted mechanisms. 
%We cannot fight with the strong theoretical results [elusive i reszta].  but still we can search for improvemenets by studying the details omitted in the derivatios

In practice the noises are  often attributed to an uncontrolled fields coupled to qubits, like electromagnetic field, which cannot be fully erased by shielding. %Can one do anything with such poorly controlled noises?
Is it possible to mitigate the effect of such phenomena? \\
It turns out that one can benefit from the effect imposed by the quantum statistics and short-range interactions, leading to the famous spin self-rephasing \cite{deutsch2010spin,buning2011extended}. It was experimentally demonstrated that this effect may increase coherence time {by orders of magnitude} of the coherent state  used in the experiment. %Can one use these techniques also in the case of the strongly correlated atomic sources, to benefit from them in interferometry? 

Here we address the question whether these techniques can be applied to the strongly correlated atomic sources (useful in the context of interferometry), trying to elucidate the role of the quantum statistics in preserving the quantum correlations in systems of interacting, trapped atoms coupled to magnetic field.

{\kp
The paper is organized as follows. In the Sec. \ref{sec:system-description} we present our model of $N$ interacting two-level atoms trapped in the harmonic trap and in the presence of the inhomogeneous magnetic field. The first results are given in  Sec. \ref{sec:Noninteracting-case}, devoted to the simplest case of the ideal gas. In this section we discuss how the quantum state loses its coherence due to the inhomogeneous magnetic field and how the process depends on the quantum statistics.
In the Sec. \ref{sec:Interacting_case} we turn to the interacting case, in which the spin self-rephasing occurs. We show how the interaction keeps the spin state completely symmetric, therefore blocking the dephasing. On the other hand, the mechanisms does not ensure that the desired properties (e.q. visibility, squeezing parameter) are preserved. This leads to the central part of the paper -- study whether the interactions can be employed to protect entanglement. We answer this question by giving an example of the squeezed states, whose squeezing is  preserved in certain window of interaction. Discussion of the results follows in Sec. \ref{sec:discussion}.
}
\section{System description.\label{sec:system-description}}
We study  a thermal cloud of $N$ bosonic two-level atoms trapped in a harmonic 1D trap, which internal degrees of freedom are spanned by two hyperfine states, which we denote by $\ket{\downarrow}$ and $\ket{\uparrow}$ (in physical realizations different atoms and internal states are used). The magnetic field is assumed to be non-uniform. 
We focus on dynamics in which an equilibrated system, a thermal cloud of $N$ atoms in $\ket{\downarrow}$ is suddenly initiated in a certain state of the internal degrees of freedom $\ket{\psi_{\text{spin}}}$, assuming unchanged spatial degrees of freedom:
\begin{equation}
\hat{\rho}(t=0) = \hat{\rho}_{\rm thermal} \otimes  \ket{\psi_{\text{spin}}}\bra{\psi_{\text{spin}}},
\label{initialstate}
\end{equation}
where $\hat{\rho}_{\rm thermal}$ is the initial spatially thermal state \ft{In the majority of our calculations, we approximated the initial thermal state with the thermal state for non-interacting atoms. We consider such low densities, that the state is close to the real thermal state of $N$ atoms in $\ket{\downarrow}$}.
Such family of states covers the typical ones, in which by $\pi/2$ pulse the atoms are in the spin coherent state with 
$\ket{\psi}$ equal to $\otimes_{i=1}^N\bb{\frac{\ket{\downarrow}_i + \ket{\uparrow}_i}{\sqrt{2}}}$ or squeezed states generated by two-axis counter twisting dynamics \cite{Kitagawa1993}.
This is standard initial state of atomic interferometers, where the fundamental quantity is the coherence
\begin{equation}
C := \left| \vec{S}\right| = \sqrt{\left< \Sx\right>^2 + \left< \Sy \right>^2},
%C := {\rm Tr} \left\{ \, | 0 \rangle\langle 1|_1  \,  \right\}.
\label{eq:contrast}
\end{equation}
useful when the dynamics takes place on the equator of Bloch sphere ($\left<S_z\right>=0$), which is the case in typical scenarios. The above is expressed with the help of the collective spin projection operators:
\begin{equation}
\Sx := \sum_{i=1}^{N}\hat{\sigma}_{x, i}\quad\quad\Sy := \sum_{i=1}^{N}\hat{\sigma}_{y, i}\quad\quad \Sz := \sum_{i=1}^{N}\hat{\sigma}_{z, i}.
\end{equation}

%{\kp  UWAGA: czy definicja koherencji spojna w pracy?}

%Due to particles indistinguishability, it is sufficient to evaluate the quantity for the first qubit, as indicated in the subscript in definition \eqref{eq:contrast}.
%The coherence is usually extracted from the interference pattern, where it corresponds to contrast. 

{\kp
The dynamics is generated by the Hamiltonian:
\begin{equation}
\hat{H} = \sum_{i=1}^N \bb{-\frac{\hbar^2}{2m}\frac{\partial^2}{\partial x_i^2}  + V(\hat x_i) + \hat{H}^{(i)}_{\rm B}}+  \sum_{i<j} \hat{V}_{\rm int}^{(i,j)}.
 \label{eq:hamiltonian}
\end{equation}
The first sum in the Hamiltonian \eqref{eq:hamiltonian} contains a sum of the single particle contributions, with a harmonic external potential $V(\hat x) = \frac{1}{2} m \omega^2 \hat x^2$ and a non-uniform magnetic field. We will show results for the case $B(x)=b_0 + b_1 x + b_2 x^2$ polarized along $Z$ direction. Therefore, the operator $\hat{H}_{\rm B}$ expressing the energy in magnetic field takes the form
\begin{equation}
\hat{H}_{\rm B}^{(i)} := \Delta\mu\, B(\hat x_i) \, \hat\sigma_{z, i},
\end{equation}
where the coefficient $\Delta\mu$ has the meaning of the differential magnetic moment between the states $\ket{\downarrow}$ and $\ket{\uparrow}$.
The second sum in \eqref{eq:hamiltonian} expresses the energy arising from  the interaction between atoms. We focus on the short-range interactions modeled by delta functions. There are three channels of such interaction, depending on the internal levels of the interacting atoms. Therefore, the interaction potential 
\begin{equation}
\hat{V}_{\rm int}^{(i,j)} := \delta(\hat x_i-\hat x_j) \bb{g_{00} \hat{P}_{i,\,j}^{00} + g_{11} \hat{P}_{i,\,j}^{11}+ g_{01} \hat{P}_{i,\,j}^{01}},
\label{eq:int-potential}
\end{equation}
depends on the projectors $\hat{P}_{i,\,j}^{\sigma\sigma'}$ which restrict the  $i$-th and $j$-th atoms to the subspace with one atom in $\ket{\sigma}$ and one atom in $\ket{\sigma'}$.
}

%OPIS DEFAZOWANIA - ROLA POLA MAGNETYCZNEGO
The position dependent magnetic field, which results in the decay of coherence, has three contributions in our model. The uniform part $b_0 \sigma_z$ only shifts the energy levels. This term alone would only rotate the whole state $\ket{\psi(t)} = e^{-i \Delta \mu\, b_0 t \hat{S}_z/\hbar} \ket{\psi(0)}$; since it commutes with every other part of hamiltonian, it may be removed by considering the state in the rotating reference frame. Because it introduces only trivial dynamics, we employ this procedure and set $b_0=0$.

 The term $b_1 x \, \hat{\sigma}_{z, i}$ shifts center of the trapping harmonic potential depending on the internal state. 
  Finally $b_2 x^2\, \hat{\sigma}_{z, i}$ modifies the trap frequency $\omega$ into {\ks$\omega' := \sqrt{\omega^2 + 2 \Delta \mu b_2 \left<\hat{\sigma}_{z}\right>}$} making it also state dependent. We restrict our consideration to small magnetic field, such that $\omega^2 \gg \Delta\mu b_2 $ -- often encountered scenario in the experiments.

\section{Noninteracting case: role of the quantum statistics\label{sec:Noninteracting-case}}
Due to the non-uniform quadratic fields one has to deal with two harmonic potentials $V_{\sigma}(x)$  dedicated to atoms in $\ket{\sigma}$, with $\sigma=\downarrow,\,\uparrow$.
We will designate eigenstates (Hermite functions) for these two potentials as $ \ket{\phi_{n, \sigma}} $. 
{\kp In the ideal gas case, the only effect of the linear term %in position of the inhomogeneus field, 
$\sum b_1 \hat x\hat \sigma_{z, i}$ is the relative shift between the eigenstates $\ket{\phi_{n, \uparrow}}$ and $\ket{\phi_{n, \downarrow}}$. In the extreme case of strong magnetic field the two harmonic traps would be separated, and then the coherence $C$ would vanish, just due to vanishing overlap between clouds of atoms in different internal states. For small magnetic field, being the subject of this paper, the linear term is less significant than the quadratic term $\sum b_2 \hat x_i^2 \sigma_{z,i}$, as discussed below. The quadratic term in the magnetic field is also an experimental issue \cite{deutsch2010spin} }.

We first show on the example of one qubit how the presence of the state-dependent potentials leads to the contrast decay in time.
For one particle, initially in a pure state $ \ket{\psi (x;\,t = 0)}\bb {\ket{\downarrow} + \ket{\uparrow}}/\sqrt{2}$, dynamics can be easily evaluated. The spatial part of the state can be expanded in both bases of eigenstates (separately for $\sigma=~\uparrow$ and $\downarrow$):
\begin{equation}
\psi (x; \, t = 0) = \sum_n c_{n, \sigma} \ket{\phi_{n, \sigma}}
\end{equation}
which leads to a formula for the state at time $t$
\begin{equation}
	\ket{\psi (t)} = e^{- iHt / \hbar} \ket{\psi (t = 0)} = \sum_n c_{n, \sigma} \, e^{- i E_{n, \sigma} t/\hbar}\, \ket{\phi_{n,\sigma}} \ket{\sigma},
	\label{eq:psiattimet}
\end{equation}
where {\kp $E_{n, \sigma} = \hbar \sqrt{\omega^2 + 2\, \Delta\, \mu\, b_2\, \langle \sigma |\hat{\sigma}_{z}|\sigma \rangle }(n+1/2)$}
%\langle \sigma |\hat{\omega'} |\sigma \rangle (n+1/2)$ 
are the eigenstates of the modified harmonic potentials and $\hat{\omega} = $
The contrast \eqref{eq:contrast} can be written as $C=|\left<S_x+i S_y\right>|$. Let us denote the term $s:=\left<S_x+i S_y\right>$; being an expectation value of single-particle operator, it is extensible -- contributions from independent atoms are simply added. Its value for a single atom in state described by Eq. \eqref{eq:psiattimet} is
\begin{equation}
s_{\text{sp}} =  \frac{1}{2} %{\rm Re}
\sum_{n, m}c_{n,0}^* c_{m,1}\,e^{- i \bb{E_{n,1} - E_{m,0}} t/\hbar} \,
\langle \phi_{n, 0} |  \phi_{m, 1}\rangle
\label{eq:contrast-sp}
\end{equation}
We consider only weak magnetic fields. Then the phase evolution will be more important than the stationary overlaps, for which we assume $\langle \phi_{n, 0} |  \phi_{m, 1}\rangle \approx \delta_{m,n}$. If the inital spatial state is $n$-th 
eigenstate of the bare harmonic potential $\ket{n}$, the value determining the constant rotates on the complex plane: ${s}_{\rm sp}^{\ket{n}} \approx\frac12 \exp \bb{-i 2 \Delta\mu b_2 n t/(m\omega)}$. In this situation the contrast is constant -- when more atoms are added, the situation is changed due to different rates of rotation.

The last formula is handy to determine the decay of the contrast. If the atom was initially in the thermal state of the bare potential,  $s^{\rm th}_{\rm sp} \approx  \frac{1-\xi}{1-\xi \, e^{-i \frac{2 \Delta\mu b_2 t}{m\omega}}} $, where $\xi = e^{-\hbar\omega/(k_{\rm B} T)}$. The computation may be put forward to a thermal state of many non-interacting atoms: the composite contrast depends on single-particle contributions weighted according to the atom occupation,
\begin{equation}
C^{\rm th}   \approx \left|\sum_{k=0}^{\infty} \frac{\langle n_k \rangle_{\rm T}}{N} s_{\rm sp}^{\ket{k}}\right|
\end{equation}
where $\langle n_k \rangle_{\rm T}$ is the thermal average occupation of the $k$th energy level.

\begin{figure}
	\includegraphics[width=0.45\textwidth]{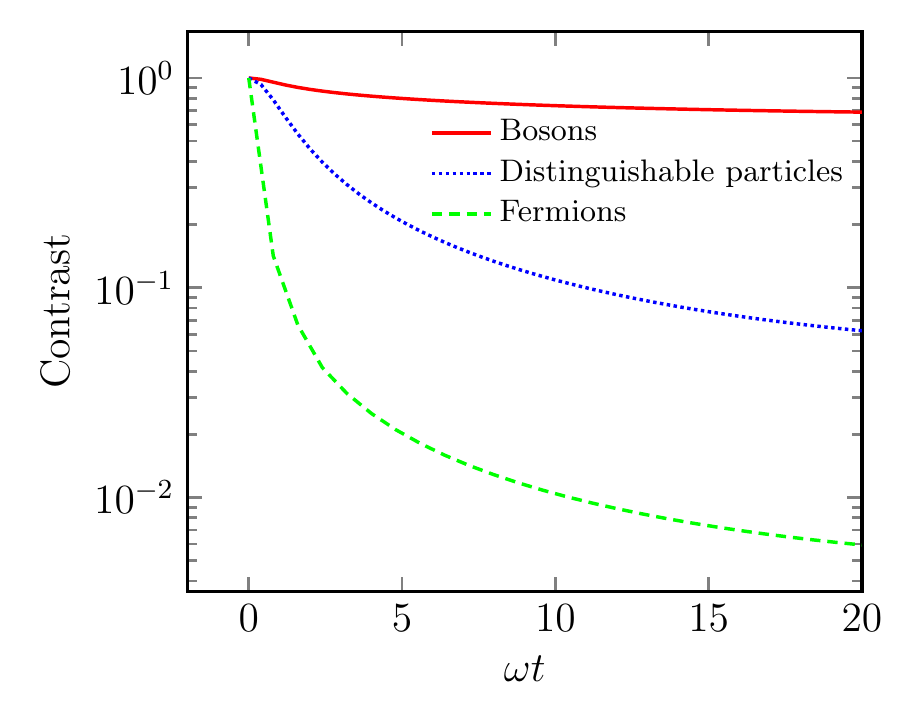}
	\caption{(color online) Decay of the coherence \eqref{eq:contrast} of ideal gas in a thermal state. The decay is due to a quadratic magnetic field with  $b_2=0.05$ (in oscillatory units), without the linear term ($b_1=0$). Temperature is $k_{\rm B}T = 10 \hbar \omega$ and number of atoms is equal to $N=100$ in all three considered cases. 
		\label{fig:ideal_statistics}}
\end{figure}

Intuitively, the bigger is the spread of atoms energies, the bigger are the differences between the phases the atoms accumulate during the evolution. By raising the gas temperature, we should observe  faster decay of the contrast. An interesting situation is when we fix a temperature and increase the number of atoms. Then one observes qualitatively completely different behaviors of the contrast decays, depending on the atomic statistics one assumes. An example for ideal gas is given in Fig. \ref{fig:ideal_statistics}, which shows the role of the atomic statistics -- a key factor in preservation of the coherence.

In the fermionic case, the thermal state in the limit $N\to\infty$ converges to a many-body state in which each single particle level up to {\ks the energy} $N\hbar\omega$ is occupied by a single atom. The spatial and momentum distributions both widen, %atomic density  spreads in both, momentum and spatial representations, 
leading to quicker dephasing and coherence decay. 
In the case of bosons the converse happens. In the limit $N\to\infty$ mostly the single-particle ground state mode becomes occupied %
-- a 1D reminiscence of Bose-Einstein condensation. The density is then dominated by  the ground state contribution, %becomes a single-particle ground state function, i.e. a Gaussian function,
 with all atoms being in a narrow region of space and momentum. Such gas dephases at a slower pace as compared to fermions and distinguishable particles.

\section{Interacting case -- spin self-rephasing.\label{sec:Interacting_case}}
%\begin{figure}
%	\includegraphics[width=0.18\textwidth]{post/many.pdf}
%	\caption{Evolution of visibility $\langle S_x\rangle$ as a function of time and number of particles $N$. Solid lines denote evolution with interaction $g=\ldots$, dashed -- noninteracting case. Temperature of spatial degrees of freedom is $T=1$. 
%		\label{fig:sxint}}
%\end{figure}

The interacting case does not admit analytical treatment, therefore numerical simulations of the dynamics have been performed.

{\kp Due to the interaction three additional parameters do appear: the interaction strengths $g_{00}$, $g_{01}$ and $g_{11}$, see Eq. \eqref{eq:int-potential}.}
%, govering the short range  interaction between cold bosons in 
To reduce the parameter space we  decided for the following  parametrization\ft{In the case of $^{87}$Rb atoms one has $c\approx 0.02g$.}:
\begin{equation}
\begin{aligned}
g_{00}&=g-c,\\
g_{01}&= g,\\
g_{11}&=g+c.
\label{eq:g-parametrization}
\end{aligned}
\end{equation}

{
As a motivation for this choice, let us mention %that the system studied in this article, the cold cloud of $N$ two-level atoms is a subject of the number 
some results in the field of Bose-Einstein condensation of cold clouds of several two-level atoms. It is known that in the single mode approximation, when all atoms occupy a single common spatial orbital, the dynamics in the internal degrees of freedom is governed by the Hamiltonian $H_{\rm OAT} = \chi\,\hat{S}_z^2 + \tilde{\chi} \hat{S}_z \hat{N}$, where $\hat N$ is a particle number operator -- a model similar to one-axis twisting hamiltonian known from \cite{Kitagawa1993}. This Hamiltonian leads to nontrivial dynamics -- a coherent state may become squeezed during the evolution. In the long time regime (not yet realized in an experiment) a Schr\"odinger cat state  -- superposition of two orthogonal spin coherent states -- appears \cite{Sorensen2001}. This entangling dynamics would make the model more complex and it would be more difficult to draw conclusions. Therefore we want a system in which the non-linear term $\chi \hat{S}_z^2$ is small, even though the interactions play an important role for spin-self-rephasing. In the aforementioned approximation the parameter by $\hat S_z^2$ vanishes when interaction strengths are parametrized as in Eq. \eqref{eq:g-parametrization}: $\chi \propto \bb{g_{00}+g_{11}-2g_{01}}$. Avoidance of the nonlinearities is the primary cause of this parametrization. % therefore we use the parametrization \eqref{eq:g-parametrization} to make the non-linearity small $\chi\approx 0$. 
%TODO wytlumacz nieliniowosc
 
In real experiment the system is quasi-1D -- the gas  is trapped in the strongly elongated harmonic trap. The corresponding effective 1D model has the coupling strength $g$ equal to  $2\hbar\omega_{\perp} a$, where $\omega_{\perp}$ is the high frequency of the transverse harmonic potential blocking the motion in the direction perpendicular to $X$-axis. 
As an example, in the case of Rubidium atoms trapped in the potential with frequencies $\bb{\omega_{x}, \,\omega_{\perp}} = 2\pi\times (17, 14000 )$ Hz, as in \cite{Meinert2015}, one has $g\approx 0.5 $ in oscillatory units based on $\omega_x$. Therefore we scan parameter space in the experimentally relevant interval $g \in [0,\,1]$.
}

In the next section we will show how the spin-self rephasing works for this parametrization and the experimentally relevant coupling strengths range.

\begin{figure}
\centering
\includegraphics[width=\linewidth]{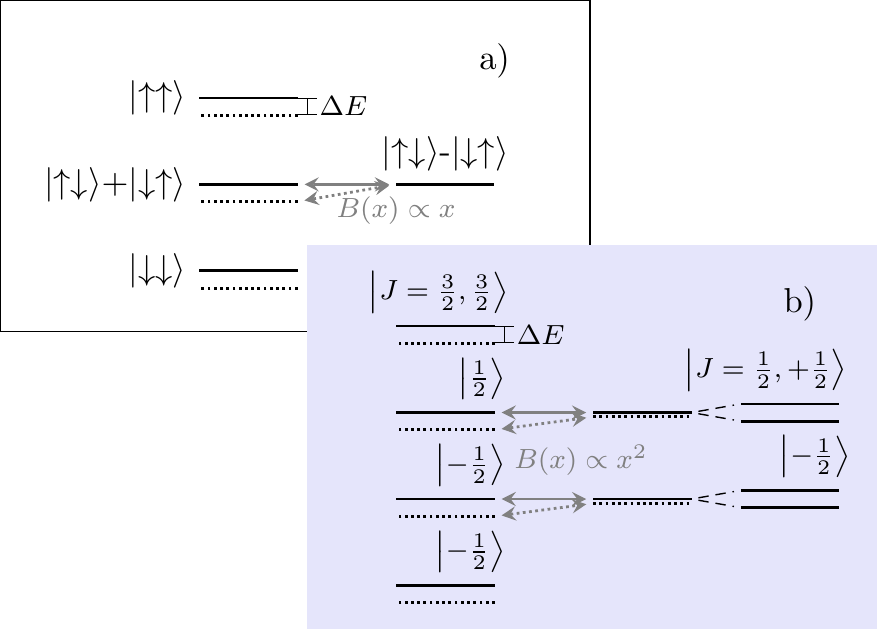}
\caption{
Energy diagram to illustrate the dephasing and the spin-self rephasing mechanisms for a) two and b) three bosons. Two (pseudo)spins can be either in the completely symmetric space (triplet) or in the singlet state; the linear magnetic field couples the two subspaces. As the subspaces have the same energies, this coupling leads to a transfer of atoms out of the symmetric subspace, i.e. to the dephasing. In the case of quadratic magnetic field, the dephasing mechanism is similar but three atoms are involved. 
Interactions between particles separate the total spin subspaces ($J=1$ and $J=0$ in a), $J=3/2$ and two copies of $J=1/2$ in b), leading to smaller effective coupling.
The figure is an adaptation of the Figure and explanation given in \cite{Gibble2010viewpoint}.
\label{fig:energy-diagram}}
\end{figure}

The spin self-rephasing phenomenon relies on the proper tuning of the interaction strengths \cite{deutsch2010spin} -- such that the energy levels shift in the symmetric subspace (see Fig. \ref{fig:energy-diagram}) overcomes the effects of the inhomogeneity of the magnetic field. This energy shift, known as the collisional shift $\delta_{\rm col}$ % \ft{This is also known as the identical spin rotation effect.},
is  approximately equal to $|g_{01}| \bar{n}$, where $\bar{n}$ is the average density of the gas. The effect of the magnetic field inhomogeneity $\Delta_B$ also introduces its own energy shift: approximately the energy originatic from the presence of the magnetic field $\langle \hat{B}(x)\rangle$. \\
Finally the so-called lateral elastic collision energy (see \cite{deutsch2010spin}), $E_{\rm lat} = 2g_{01}^2 \bar{n}v_{\rm T}/(3\pi\sqrt{\pi} m)$  should be smaller than $\hbar \omega $ (Knudsen regime) and the collisional energy $\delta_{\rm col}$. 
We introduced $v_{\rm T}$ which is the thermal velocity equal to $\sqrt{k_{\rm B}T/m}$; $T$ is the temperature.
In the simulations presented in this paper, $\Delta_B$ and $E_{\rm lat}$ are of the order of $10^{-2}$ and $\delta_{\rm col}$ is of the order of $1$ in oscillatory units.

%TODO rysunki dla dwoch i trzech atomow nakladaja sie na siebie

\subsection{Coherent states}
In this section the state ($\ket{\psi_{\text{spin}}}$ in Eq. \eqref{initialstate}) is initiated as maximal spin projection of $S_x$; the aim is to tune interactions in such a way that the coherence $C$ (Eq. \eqref{eq:contrast}) decays slower. %Since no constant clock term $\propto S_z$ is present in the Hamiltonian, the state should not rotate on the composite Bloch sphere and the main observable of interest is $S_x$. --- 

{\kp It has been experimentally demonstrated that the dephasing can be substantially blocked thanks to the interaction between atoms in the so called spin-self-rephasing effect \cite{deutsch2010spin, buning2011extended, Solaro2016, Fuchs2002, Lhuillier1985}.}
In articles following the initial discovery of spin self-rephasing importance of total spin preservation has been studied as well \cite{deutsch2010spin}. The original explanation {\kp is illustrated by the diagrams in {\ks Fig. \ref{fig:energy-diagram}a)} } involved analysis of two-particle subsystems: initially, the particles are in triplet spin state; inhomogenous magnetic field couples $\ket{j=1; m=0}:=\left(\left|\uparrow\downarrow\right>+\left|\downarrow\uparrow\right>\right)/\sqrt{2}$ to the antisymmetric singlet spin state $\ket{j=0; m=0}=\left(\left|\uparrow\downarrow\right>-\left|\downarrow\uparrow\right>\right)/\sqrt{2}$, effectively lowering population of maximal total angular momentum. 
{\kp 
As we deal with bosons, the atoms can be in the singlet states in the internal degrees of freedom only if simultaneously their spatial wavefunction is antisymmetric. The interaction energy coming from the short-range potential is $0$ in the singlet subspace -- a result of spatial wavefunction being antisymmetric as well. On the other hand if the internal state of atoms is in the symmetric triplet manifold, then also their spatial wavefuction has to be symmetric, and these atoms do interact. 
}
Therefore, in the case of interacting bosons energy levels of triplet spin substates are shifted {\kp by $\Delta E$ (see Fig. \ref{fig:energy-diagram})}, but the energy of singlet case is constant (the spatial state is antisymmetric). As a result, population transfer is diminished. 
{\kp In the case of the magnetic field changing quadratically in space the dephasing mechanisms is slightly more complicated, as there are at least three atoms needed to demonstrate the population transfer between total spin subspaces, as shown in Fig. \ref{fig:energy-diagram}b).
}

To visualize {\kp the above-mentioned transfers of population}, let us introduce the spin density matrix of a system: after partial trace of spatial degrees of freedom, only internal degrees of freedom are left. We can rearrange them to form the block diagonal spin density matrix, containing all of the information useful in quantum metrology:
\begin{equation}
\hat\rho_{\text{spin}}=\begin{pmatrix}
\hat\rho_{j=N/2}& &\ldots&\\
& \hat\rho_{j=(N-1)/2} &  &\vdots\\
\vdots & & \ddots & \vdots\\
&\ldots&&\hat\rho_{j=j_{\text{min}}}\end{pmatrix},
\end{equation}
where $j_{\text{min}}=1/2$ for odd $N$, otherwise $j_{\text{min}}=0$. 
{\kp The subspace with the maximal collective (pseudo)spin, $j=N/2$ corresponds to completely symmetric states. }
Usually in metrological scenarios the state is initialized so that only $\rho_{j=N/2}$ part is populated (it is the case for coherent and squeezed states).
%and other nonstandard states (e.g. Schrödinger cat, being the superposition of coherent states). 
Population conservation of this part of spin density matrix is hence a natural goal.

\begin{figure}[!h]
\centering
\includegraphics[width=.9\linewidth]{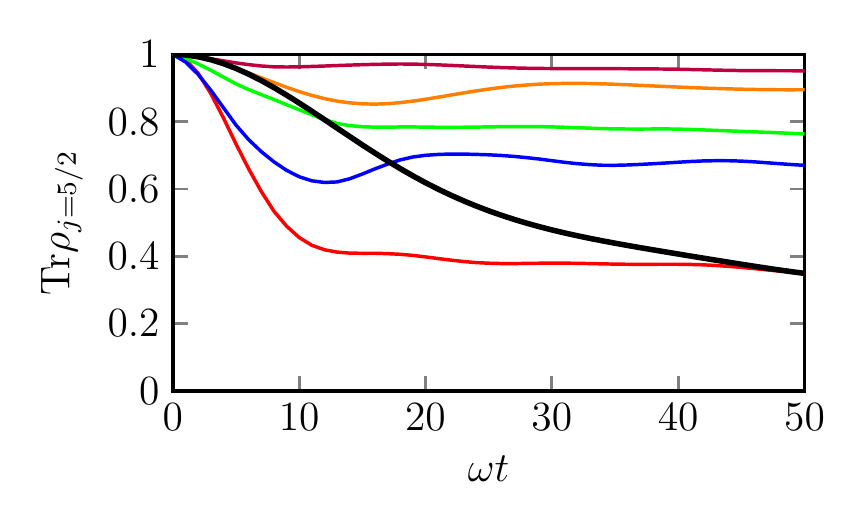}\\
\includegraphics[width=.9\linewidth]{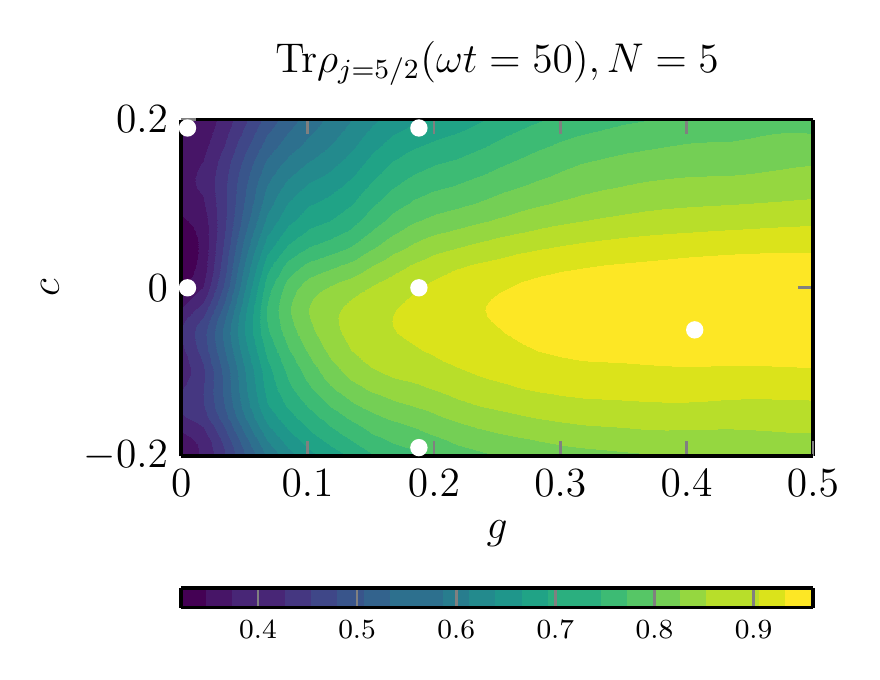}
\caption{Top: Evolution of maximal spin population as a function of time for several interaction strengths denoted in the bottom image by dots (the noninteracting case is denoted by thick black line), $N=5, k_{\rm B}T=3\hbar\omega$ and quadratic inhomogeneity $b_2 = 0.01 $. Quick decay followed by plateau is visible.  \\
Bottom: Preservation of trace for the same parameters for time $\omega t=50$ as a function of interactions parametrized by mean interaction strength $g$ and $c$.\label{fig:symmetric}}
\end{figure}

{\kp As demonstrated in Fig. \ref{fig:symmetric},} in the simulations, the desired behavior is observed: when interactions are present, a large amount of maximal spin population is preserved. 
{\kp the number of simulations is summerized in the lower panel of  Fig. \ref{fig:symmetric}, showing the map of weights of the maximal spin subspace at fixed time $\omega t = 30$.
By this metric, total spin is preserved well if high enough interaction between particles is present; wide range of interaction strengths works well.
%It seems that the spin-self rephasing works for any finte interaction strength $g$.
}

%{\kp O co chodzi w tym zdaniu? While preservation of spin states may require careful tuning of interactions strengths, the total spin is conserved if mean interaction is high enough.}

\begin{figure}[!h]
\centering
\includegraphics[width=\linewidth]{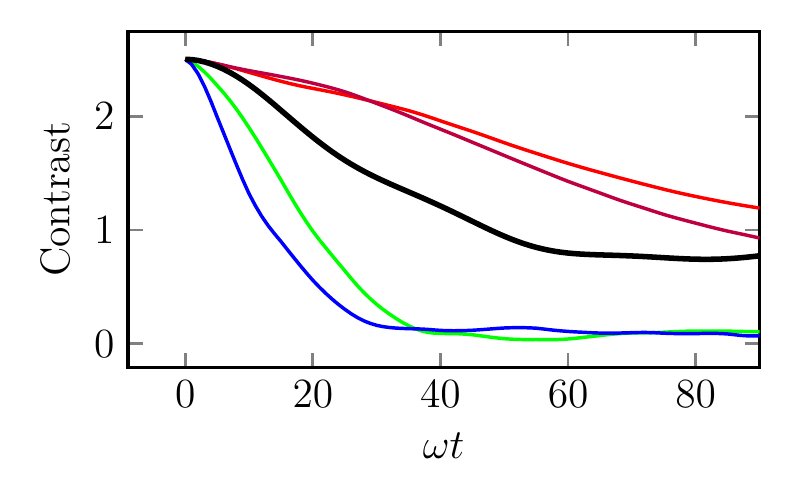}
\includegraphics[width=\linewidth]{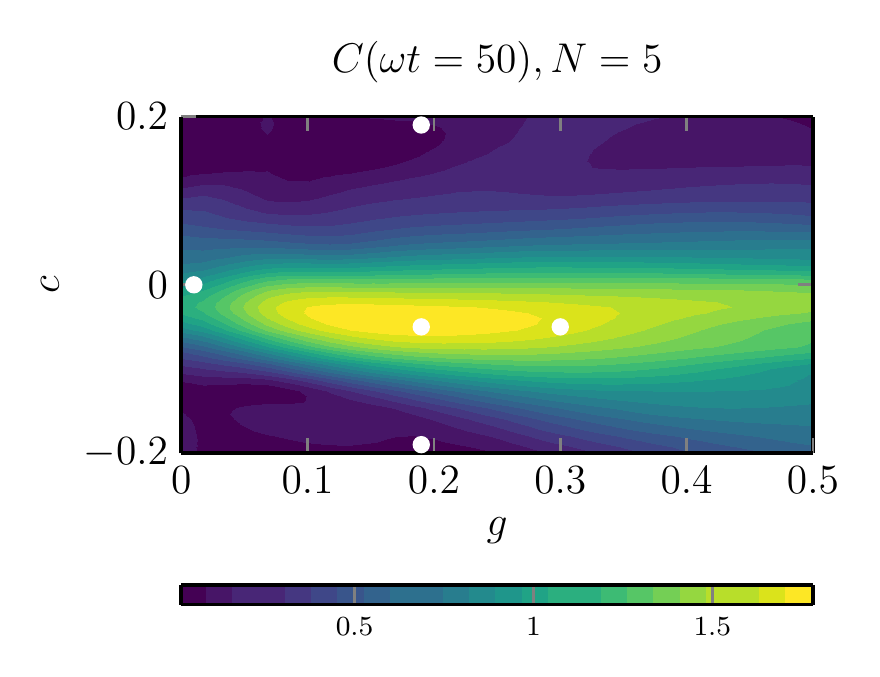}
\caption{Effect of interactions on preservation of the coherence $C$. Total number of particles $N=5$, the temperature $k_{\rm B} T=3\hbar \omega$, quadratic inhomogeneity $b_2=0.01$. Top:  
typical short time evolution of $C$ for several interaction strengths denoted in the bottom image by dots, the noninteracting case is denoted by thick black line. As evidenced by the Figure, it is possible to approximately double the time of decoherence (when the contrast attains half of its original value).
Bottom: $\langle S_x\rangle$ at $\omega t=50$ for various values of interaction strengths parametrized by $g$ and $c$. Please note that the same interaction range is used in Fig. \ref{fig:symmetric}.\label{fig:Sx}}
\end{figure}

{\kp As shown above, interactions help in preservation of the total spin. To determine how the coherence is protected in the same scenario, Fig. \ref{fig:Sx} shows the its dynamics and the map of values of $C$ at fixed instant of time $\omega t=50$ in the broad range of interaction strengths. 
Although the state remained symmetric, apparently its dynamics is quite complex, and in many cases its coherence has been lost. This is caused in part by the small transfer between the symmetric  and the other subspaces, but also the interactions may separate the clouds of atoms in different internal states (spin-position entanglement) or introduce strong, here unwanted, quantum correlations.

A natural question arises whether it is possible to use the spin-self-rephasing to prevent not only the first order-coherence --  already shown in the experiment -- but also more subtle correlations. The simulations performed do not indicate usefulness in protection of the highly entangled states like the Schr\"odinger cat state ($\ket{N0}+\ket{0N}$); from now on we focus on squeezed states of practical importance in the quantum metrology.
}

\subsection{Squeezed spin states}
Typical preparation of squeezed spin state involves utilization of particle-particle interactions to produce effective Hamiltonian quadratic in spin components. Few of the examples are: one axis twisting $S_z^2$ {\kp (implemented in the number of experiments \cite{esteve2008, Gross2010,SchleierSmith:2010eo, Riedel2010})} or two-axis counter-twisting (TACT) $S_y S_z+S_z S_y$. Spin squeezed states are produced as a result of the evolution. Here we use the TACT model of spin squeezing, in which the initial squeezed state is generated by evolution of coherent state $\left|+\right>$ (the eigenstate to maximal $S_x$ eigenvalue):
\begin{equation}
\left|\psi_\theta\right> = \exp\left(i \theta (S_y S_z+S_z S_y)\right) \left| + \right>.
\label{eq:ssstate}
\end{equation}
The TACT Hamiltonian is used only in the preparation of the spin state for further numerical evolution according to Hamiltonian Eq. \eqref{eq:hamiltonian}. During the preparation only spin degrees of freedom are taken into account (in particular, inhomogenous magnetic field does not affect the initial state).
Squeezing parameter quantifies the usefulness of the state in quantum metrology, as compared with the spin coherent states \cite{Wineland1994}. The definition depend on the choice of direction of spin observation $\vec n$ and a perpendicular direction $\vec n_\perp$:
\begin{equation}
\xi^2_\psi = \frac{N \left(\Delta^2 \vec S\cdot \vec n_\perp\right)_\psi}{ \left<\vec S\cdot \vec n\right>_\psi^2}.
\label{eq:xi}
\end{equation}
In our calculations we use $\vec n\propto \left<\vec S\right>$. Since $\left<S_z\right>=0$ during the evolution and the spin dynamics is confined to equator of the Bloch sphere, we choose $\vec n_\perp=(n_y,-n_x,0)$. This variable is a measure of entanglement between individual spins: if $\xi^2<1$, the spins are entangled.
In the following paragraphs we study the dynamics of the squeezing parameters, assuming the initial spin state prepared by TACT Hamiltonian (Eq. \eqref{eq:ssstate}).

\begin{figure}[!h]
\centering

\includegraphics[width=\linewidth]{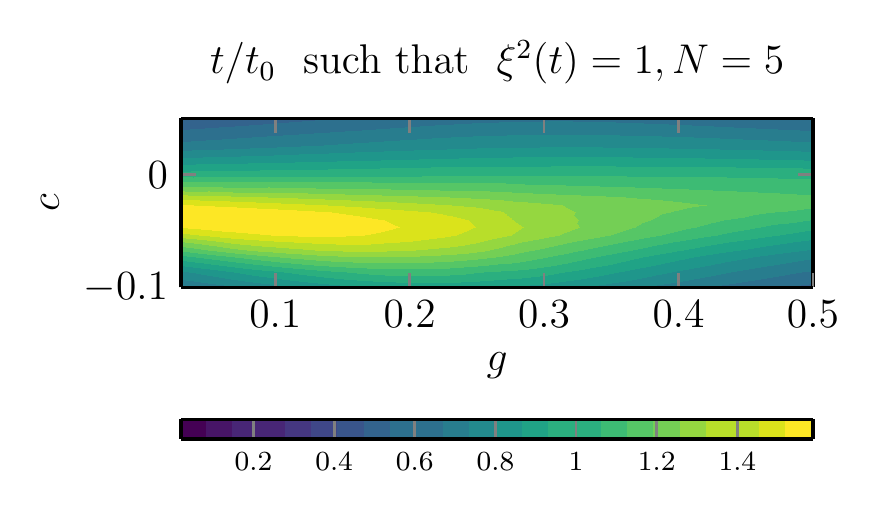}\\
\includegraphics[width=\linewidth]{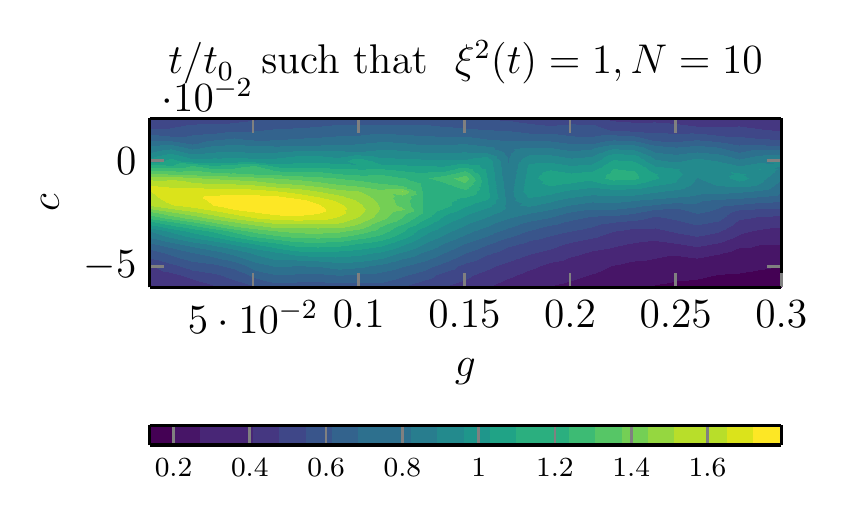}

\caption{Effect of interaction on preservation of squeezing parameter $\xi^2$ for various values of interaction strengths parametrized by $g$ and $c$. The color is correlated with the time of decoherence $t$ at which squeezing parameter $\xi^2$ becomes larger than 1 as compared to the time of decoherence without interactions $t_0$. The temperature $k_{\rm B} T=3\hbar \omega$, quadratic inhomogeneity $b_2=0.01$; total number of particles: top: $N=5$, bottom: $N=10$. In both cases the squeezing time ($\theta$ in Eq. \eqref{eq:ssstate}) is $0.05$, leading to $\xi^2(N=10)=0.44$ and $\xi^2(N=5)=0.69$ at the beginning of the evolution.\\
 In the case of $N=10$ it is possible to increase the time at which $\xi^2$ becomes higher than $1$ by a factor of $75\%$, for $N=5$ this time can be increased by about $60\%$.
 \label{fig:squeezing}}
\end{figure}

Our results are shown in Fig. \ref{fig:squeezing}. For 10 particles even for relatively long evolution time, $\omega t = 30$, one can restore squeezing $\xi<1$ in  certain windows of parameters, which corresponds to delay of the decoherence time of about 75\% (for $N=5$ the decoherence time can be delayed by about 60\%). The large interaction strengths are not desired:  they are outside the parameter range required by spin-self rephasing (a result of $E_{\rm lat}$ increase) and induce strong correlations along with complicated spatial dynamics.

Our results indicate that the protection of squeezed states is lower as compared to coherent states, however still reasonably large to expect good results when number of particles is higher. Schr\"odinger cat states -- superpositions of two coherent states -- on the other hand are known to be extremely fragile to any source of noise, which is reflected in our simulations: their coherence is quickly lost and particle-particle interaction offers no protection.

%In the case of $N=5$ it is possible to delay the decoherence time by $60\%$

\section{Discussion and conclusions\label{sec:discussion}}

{\kp
In this paper we investigated {\it ab initio} a system of several two-level particles (pseudo-spins) in one-dimensional harmonic potential. In such an approach, all effects typically considered in metrology in an approximated way, such as collisional shift, noises, or identical spin-self rotation effect, are automatically and rigorously taken into account, as incorporated in the dynamics generated by a many-body Hamiltonian. 

The subject of the paper is to show that the spin-self-rephasing mechanism can be used to prevent decoherence of the entangled states induced by a inhomogeneous magnetic field. In the semiclassical description, each pseudo-spin would rotate around magnetic field polarization axis, with an angular velocity depending on the value of the local field $B(x)$. As a result, the collective spin decays in time. From one point of view, the dephasing is caused by populating a space with a total spin of less than $N/2$. This unwanted process can be inhibited by tuning the interactions between particles: in the interacting system, the subspaces with different total spin have different interaction energies, which stops the population transfer between them
\cite{deutsch2010spin}. 
On the other hand, even if the state remains completely symmetric (total spin $j=N/2$), the interactions also contribute to the different type of dephasing, as shown in Figure \ref{fig:Sx}. One such effect is the introduction of collisional shift, and in the case of larger interactions, even non-linear effects leading to phase collapse. 
In the light of emerging quantum technologies, it is not clear whether the interactions can be used to protect entanglement. In this article we have shown that there is a window of parameters in which the squeezed state obtained from the TACT model is stable despite the presence of nonlinear magnetic field. In the example for $10$ particles, the time scale at which the squeezing disappears can be extended by a factor of $75\%$ using reasonably low interaction strengths. 

}

Financial support by the Ministry of Science and Higher Education Grant No.0273/DIA/2016/45 is gratefully acknowledged.
\appendix

\section*{Appendix: Simulation method}
Our implementation provides a way to determine the evolution of state vectors and expectation values of observables of a system consisting of several bosons in weakly inhomogenous harmonic trap, both in high and low temperature limit. The numerical simulations assume unitary evolution generated by Hamiltonians composed of
\begin{enumerate}
\item pseudospin components, $S_x, S_y, S_z$, and polynomials of them, $S_z^2, S_x S_y+S_y S_x,\ldots$,
\item spin-position couplings, emerging from the trap inhomogenities: $ \sum \hat x_i \, \hat\sigma_{z, i}, \sum \hat x_i^2 \, \hat\sigma_{z, i}, \ldots$
\item particle-particle interaction (Eq. \eqref{eq:int-potential}).
\end{enumerate}

In high temperature scenario we use a Quantum Monte Carlo technique, which requires construction of sub-basis around fixed spatial Fock state, which corresponds to the subspace traversed by the state under time evolution.

\subsection{Initial state}
The initial state is assumed to be thermal in spatial degrees of freedom and pure in spin: the spin part is in fact assumed to be in the minimal spin projection $S_z$ eigenstate. The resulting state is subsequently {\kp rotated} by $S_y$, so that it is now in maximal spin projection of $S_x$, with the spatial degrees of freedom left intact; if needed, the state can also be squeezed by TACT Hamiltonian. This is motivated by usual experimental realizations, in which trapped atoms are selected by their internal state and left to thermalize in the trap, then a sequence of pulses is applied to the trapped ensemble and later evolution is studied.
\subsection{Thermal state}
The initial state thus depends on the properties of Hamiltonian restricted to the subspace with minimal spin projection, $S_z=-N/2$. In low temperature limit, the relevant part of the Hamiltonian can be diagonalized and the eigenstates evolved according to the procedure described above with appropriate weights.\\
The simulation of high temperature states is more intricate: relevant part of the Hamiltonian (i.e. the subspace in which majority of the population resides) has high enough dimension so that no direct diagonalization might be applied. There is however a method of random sampling from the thermal density matrix, often called Quantum Monte Carlo, which greatly reduces the computational cost needed to determine the evolution of initially thermal state at the expense of deterministic behavior. Consider the thermal density matrix,
\begin{equation}
\rho_T = \frac{\exp(-H/T)}{\Tr \exp(-H/T)}.
\end{equation}
By denoting $\Tr \exp(-H/T)$ by $Z$, it is possible to rewrite the above expression by inserting identity matrix:
\begin{equation}
\rho_T = Z^{-1} \exp\left(-\frac{H}{2T}\right)\left[\sum_i \ket{i}\bra{i}\right]\exp\left(-\frac{H}{2T}\right).
\end{equation}
Sampling from the density matrix can now be understood as approximating the identity matrix with finite number of elements $\ket{i}\bra{i}$ -- this is the same as choosing random vectors from a (fixed) basis $S$ with uniform probability and evolving them by $\exp\left(-\frac{H}{2T}\right)$. This of course works the best if $S$ is as close to the eigenbasis of $H$ as possible. In the simulations we have chosen $S$ to be the eigenbasis of harmonic oscillator without inhomogeneities and interactions -- since both of these terms of Hamiltonian are small as compared to the bare harmonic oscillator energy, the assumption is fulfilled. 

The sampled vectors are not normalized, since $\exp\left(-\frac{H}{2T}\right)$ is not unitary. This is anticipated behavior: length of the vector indicates the relative weight when the sampled ensemble is brought together in order to calculate the expectation values. Samples can be further evolved unitarily to determine the dynamics of a system, and this is the procedure employed in the simulations.  In our case, samples are drawn until the appropriate expectation values in the relevant time span converge to a stable value.  
%In Fig. \ref{fig:ideal_statistics} we show the evolution of the contrast in the ideal gas. As one can decay of contrast of bosons is much slower than for the distinguishable atoms, which in turn are more robust than fermions. Our understanding is following. At any finite temperature
\section*{Appendix: Importance of spatial dynamics}
\begin{figure}[!h]
\centering
\includegraphics[width=\linewidth]{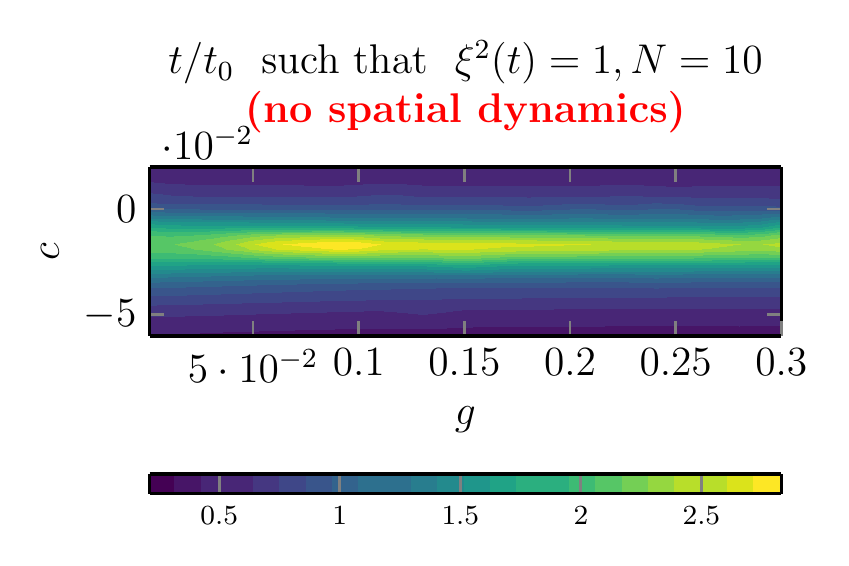}

\caption{Figure analogous to bottom picture of Fig. \ref{fig:squeezing} (preservation of squeezing parameter for $N=10$), note the same range of interaction parameters. In order to prepare this picture, the simulation basis has been artificially restricted, effectively leading to ignoring of spatial dynamics. The oversimplified description leads to decoherence time delay of about 180\%, far from 75\% obtained with more accurate simulation. \label{fig:comp}}
\end{figure}

Proper treatment of spatial degrees of freedom is important in order to obtain meaningful result, especially when large interaction strengths are present. To determine the effect of spatial dynamics, we have performed additional simulations with artificially restricted basis. The restriction effectively reduces the effect of particle-particle interactions to energy shift, which is not an accurate description for high interaction strengths. This is illustrated in Fig. \ref{fig:comp}, in which results analogous to these in Fig. \ref{fig:squeezing} are presented. The discrepancy between the two pictures is clear and demonstrates importance of full quantum description in analysis presented in this paper. \\
However, such simplified simulations are convenient in analysis of noninteracting bosons. Provided the magnetic field is small, little spatial dynamics is introduced, the description is sufficient and useful in calculation of e.g. evolution of squeezing parameter (Eq.  \eqref{eq:xi}), which depends on expectation values of two-particle operators though variance.
\bibliographystyle{plain}
\bibliographystyle{apsrev4-1}
\bibliography{refs}
\end{document}